 \documentclass[smallabstract,smallcaptions]{dccpaper}

\usepackage{epsfig}
\usepackage{citesort}
\usepackage{amsmath}
\usepackage{amssymb}
\usepackage{color}
\usepackage{url}
\usepackage{adjustbox}
\usepackage{enumitem}
\usepackage{algorithm}
\usepackage{algorithmic}
\usepackage{xcolor}
\usepackage{subcaption}

\newlength{\figurewidth}
\newlength{\smallfigurewidth}
\begin{document}

\title
{\large
\textbf{Bidirectional Learned Facial Animation Codec for Low Bitrate Talking Head Videos}
}

\author{%
Riku Takahashi, Ryugo Morita, Fuma Kimishima, Kosuke Iwama and Jinjia Zhou\\[0.5em]
{\small\begin{minipage}{\linewidth}\begin{center}
\begin{tabular}{c}
Hosei University, Tokyo, Japan\\
\end{tabular}
\end{center}\end{minipage}}
}

\maketitle
\thispagestyle{empty}

\begin{abstract}
Existing deep facial animation coding techniques efficiently compress talking head videos by applying deep generative models.
Instead of compressing the entire video sequence, these methods focus on compressing only the keyframe and the keypoints of non-keyframes (target frames).
The target frames are then reconstructed by utilizing a single keyframe, and the keypoints of the target frame. 
Although these unidirectional methods can reduce the bitrate, 
they rely on a single keyframe and often struggle to capture large head movements accurately, resulting in distortions in the facial region.
In this paper, we propose a novel bidirectional learned animation codec that generates natural facial videos using past and future keyframes.
First, in the Bidirectional Reference-Guided Auxiliary Stream Enhancement (BRG-ASE) process, we introduce a compact auxiliary stream for non-keyframes, which is enhanced by adaptively selecting one of two keyframes (past and future). This stream improves video quality with a slight increase in bitrate.
Then, in the Bidirectional Reference-Guided Video Reconstruction (BRG-VRec) process, we animate the adaptively selected keyframe and reconstruct the target frame using both the animated keyframe and the auxiliary frame.
Extensive experiments demonstrate a 55\% bitrate reduction compared to the latest animation based video codec, and a 35\% bitrate reduction compared to the latest video coding standard, Versatile Video Coding (VVC) on a talking head video dataset. It showcases the efficiency of our approach in improving video quality while simultaneously decreasing bitrate.
\end{abstract}

\section{INTRODUCTION}

\label{sec:intro}
In the digital era, video media has become a widely utilized application, with its prevalence relying on the advancement and dissemination of video coding technologies. 
Recent work on deep learning-based video compression methods has shown the possibility of compressing talking head videos at a very low bitrate for videoconferencing applications. 
DAC \cite{konuko2021ultra} proposes a deep learning approach for ultra-low bitrate video compression that reconstructs target frames by using a keyframe and target keypoints representing facial features.
HDAC \cite{konuko2022hybrid} uses non-key frames with a high compression ratio as auxiliary information to handle background motion and disocclusions that cannot be captured by keypoints. The limitations of keypoints are overcome by using animated and auxiliary frame features to reconstruct the target frame.
RDAC \cite{konuko2023predictive} uses the residual difference between the animated target frame and the target frame to correct the animated target frame. 
Furthermore, by efficiently encoding the residuals between consecutive frames over time, temporal dependencies are eliminated, leading to improved compression efficiency.
With advanced image animation techniques \cite{Siarohin_2019_NeurIPS,siarohin2019animating}, these animation based video compression methods \cite{konuko2021ultra,konuko2022hybrid,konuko2023predictive} have demonstrated superior results.

However, these unidirectional methods rely on a single keyframe, and as the target frame moves away from the keyframe, the loss of temporal correlation makes it difficult to capture large facial movements, leading to distortions in the face region of the generated frames. Additionally, the bit costs associated with ancillary information are substantial and affect overall bitrate efficiency.

\begin{figure}[t]
\centering
\includegraphics[width=0.9\textwidth]{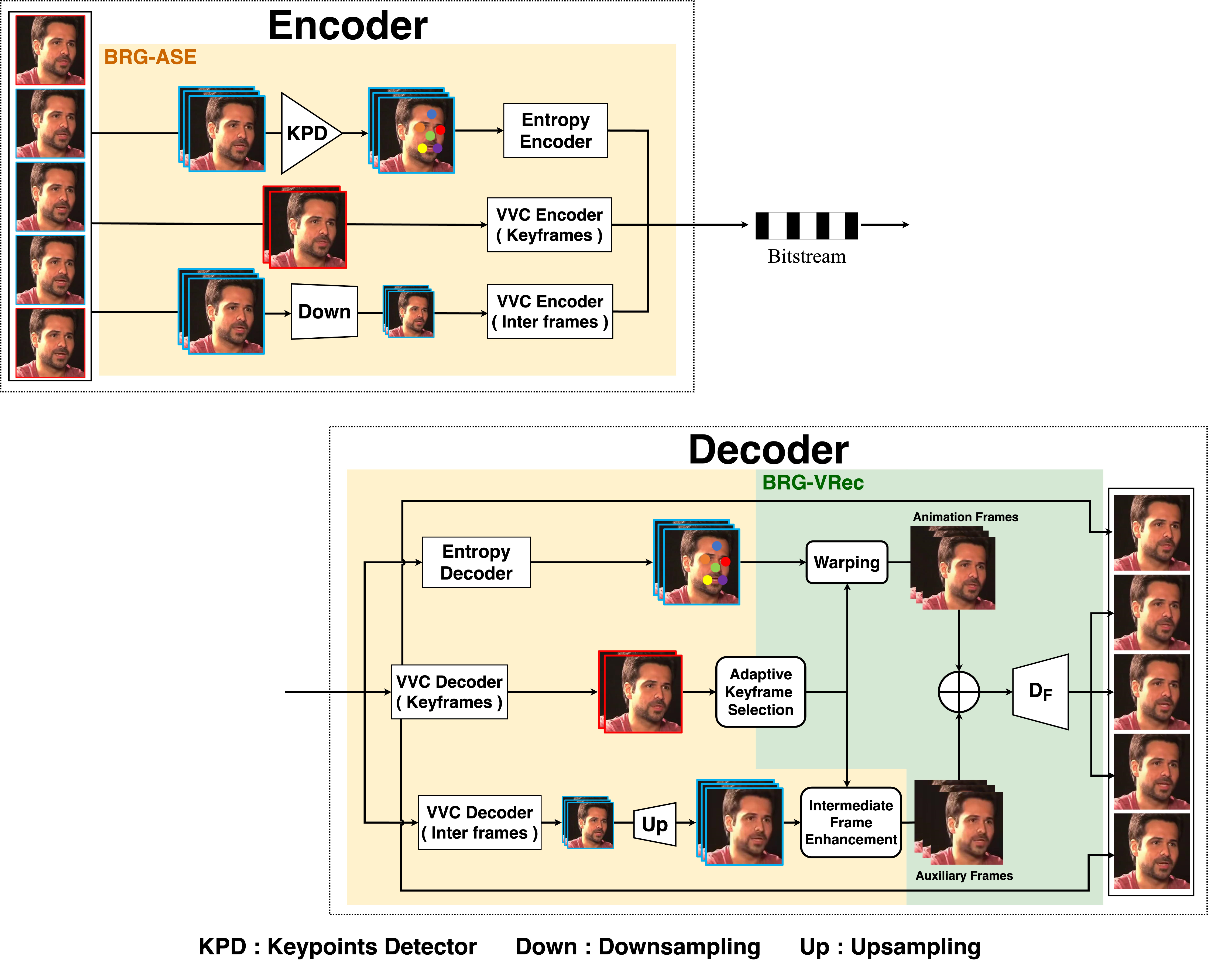}
\caption{Architecture of the proposed codec. The first and last frames (keyframes) are enclosed in red boxes, while the intermediate frames are enclosed in blue boxes. BRG-ASE stands for the Bidirectional Reference-Guided Auxiliary Stream Enhancement process, while BRG-VRec stands for the Bidirectional Reference-Guided Video Reconstruction process.}\label{fig:fig1}
\end{figure}

To solve these problems, we propose a novel bidirectional learned video codec. This codec combines an image animation method with downsampling based video coding, providing bidirectional keyframe information and low bitrate auxiliary information.
We introduce a novel coding approach based on Group of Pictures (GOP) that utilizes bidirectional keyframes (both past and future). This approach improves the quality of the video reconstruction without increasing the bitrate. 
We also introduce an adaptive keyframe selection algorithm that dynamically selects the keyframe with the highest similarity to the target frame from the bidirectional keyframes. This algorithm captures the motion between frames and prevents the loss of temporal correlation, leading to a more faithful reconstruction of the target frame.
Additionally, a lightweight auxiliary stream for non-keyframes is incorporated. By enhancing this compact data, it serves as supplementary reference information to further boost video quality with only a slight increase in bitrate. 



Specifically, our framework can be divided into two processes. In the Bidirectional Reference-Guided Auxiliary Stream Enhancement (BRG-ASE) process, we provide low-bit auxiliary frames. The downsampled frames at the encoder are transmitted, upsampled at the decoder, and enhanced using the keyframes selected by the adaptive keyframe selection algorithm from two keyframes. These frames are then used as auxiliary frames in the next process to reconstruct the target frame. This approach allows us to deliver auxiliary information while maintaining quality and avoiding an increase in bitrate.
In the Bidirectional Reference-Guided Video Reconstruction (BRG-VRec) process, the keyframes selected by the adaptive keyframe selection algorithm are animated, and we reconstruct the target frame using the animated keyframe along with the auxiliary frame obtained from the BRG-ASE process. By using keyframes selected to minimize the motion difference with the target frames, we address the issue of facial distortion caused by loss of temporal correlation.

We demonstrated an average bitrate reduction of 24\% compared to HDAC \cite{konuko2022hybrid}, 55\% compared to the latest animation based codec RDAC \cite{konuko2023predictive}. 
Additionally, we achieved a 35\% reduction compared to the low-delay configuration of the latest video coding standard, Versatile Video Coding (VVC) \cite{bross2021overview}.  
This demonstrates the effectiveness of our approach in reducing bitrate while maintaining high video quality.

\begin{figure}[t]
\centering
\includegraphics[width=0.9\textwidth]{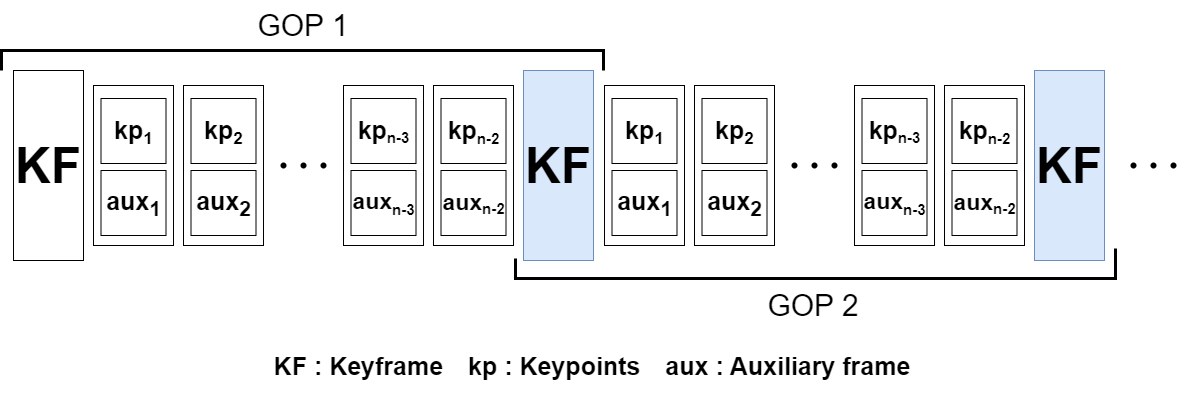}
\caption{The composition of the group being processed. The keyframe and the auxiliary frame are the same size, 256x256. The last frame (filled with blue) of each GOP is temporarily stored in the decoder for use in the next GOP. Therefore, there is no need to send the same keyframe twice.}\label{fig:fig2}
\end{figure}

\section{THE PROPOSED METHOD}
\subsection{Framework of bidirectional learned facial amination codec}

The structure of our proposed codec is shown in Figure \ref{fig:fig1}. Our codec consists of two processes: Bidirectional Reference-Guided Auxiliary Stream Enhancement (BRG-ASE; Sec \ref{sec:BRG-ASE}) and Bidirectional Reference-Guided video Reconstruction(BRG-VRec; Sec \ref{sec:BRG-VRec}).  We divide one video sequence into $N$ frames defined as the GOP $\{ {I_1, I_2, \ldots, I_{N-1}, I_N} \}$.
 

In the BRG-ASE process, at the encoder, we estimate the keypoints from the intermediate frames (intermediate keypoints), encodes them using a standard arithmetic coding for entropy coding.
Additionally, the first and last keyframes $\{I_1,I_N\}$ are compressed with high quality, while the intermediate frames 
$\{{I_2, \ldots, I_{N-1}}\}$ are downsampled and compressed.
Afterward, these compressed keypoints and frames are transmitted as a bitstream.
We note that the last keyframe (future keyframe) in one GOP is reused for the first keyframe (past keyframe) in the next GOP, as shown in Figure \ref{fig:fig2}. 
At the decoder, the intermediate frames are upsampled and enhanced by utilizing a keyframe adaptively selected from two high-quality keyframes.
The enhanced intermediate frames are used as auxiliary frames during the BRG-VRec process to address large facial movements and object occlusions that cannot be handled solely by keypoints.

In the BRG-VRec process, we generate animation frames using adaptively selected keyframes and intermediate keypoints. We then combine the features of animation frames and the auxiliary frames by the BRG-ASE process and reconstruct the frames by the decoder $D_F$. 

\begin{figure}[t]
\centering
\begin{minipage}{0.9\textwidth}
\begin{algorithm}[H]
    \caption{Adaptive frame enhancement \& reconstruction}
    \label{alg1}
    \begin{algorithmic}[1]    
    \REQUIRE $key_1,key_2$(keyframes), $auxs$(auxiliary frames),\\ 
             $inter\_kps$(keypoints extracted from intermediate frames)

    \STATE $Save(key_2)$ \COMMENT {\textcolor{blue}{Save keyframe at the decoder for next gop}}
    \STATE $key\_kp_1 = Keypoint(key_1)$
    \STATE $key\_kp_2 = Keypoint(key_2)$
    
    \FOR{$aux,inter\_kp$ $\textbf{in}$ $zip(auxs,inter\_kps)$}
    \IF{$PSNR(key_1,aux)>PSNR(key_2,aux)$}
    \STATE $enh\_aux = Enhance(key_1,aux)$ 
    \STATE $animation = Warp(key_1,key\_kp_1,inter\_kp)$
    \STATE ${DECI} = D_F(Concat(animation,enh\_aux))$
    \ELSE
    \STATE $enh\_aux = Enhance(key_2,aux)$
    \STATE $animation = Warp(key_2,key\_kp_2,inter\_kp)$
    \STATE ${DECI} = D_F(Concat(animation,enh\_aux))$
    \ENDIF
    \ENDFOR
    
    \end{algorithmic}
\end{algorithm}
\end{minipage}
\end{figure}

\subsection{Bidirectional Reference-Guided Auxiliary Stream Enhancement (BRG-ASE)}
\label{sec:BRG-ASE}

At the encoder, we estimate the keypoints $\boldsymbol{inter\_kps}$ from the intermediate frames, encodes them using a standard arithmetic coding for entropy coding, and transmits them as a bitstream. 
Following to \cite{konuko2022hybrid}, we use the U-Net \cite{ronneberger2015u} to predict 10 keypoints for facial features. 
We note that the keypoints of the keyframes are predicted at the decoder.
Furthermore, the first keyframe $\boldsymbol{key_1}$ and last keyframe $\boldsymbol{key_2}$ are compressed using VVC in the intra configuration and sent as a bitstream. 
The intermediate frames are downsampled with a scale of 2. The downsampled frames are compressed using VVC in the low-delay configuration and also sent as a bitstream. 
At the decoder, the downsampled intermediate frames are then upsampled with a scale of 2 using bicubic technique and denoted as $\boldsymbol{auxs}$. Then the $auxs$ are enhanced with the keyframes $\{{key}_1,{key}_2\}$. These enhanced intermediate frames $\boldsymbol{enh\_auxs}$ are then used as auxiliary frames to improve the accuracy of the reconstruction in the BRG-VRec process.

As shown in Algorithm\ref{alg1}, when performing the enhancement, one frame is selected from two decoded high-quality keyframes as the reference frame. 
PSNR1 and PSNR2 are calculated using the upsampled intermediate frame $aux_i \in auxs$ with ${key_1}$ and ${key_2}$, respectively.
PSNR1 and PSNR2 are compared, and the $aux_i$ frame is enhanced using ${key_1}$ or ${key_2}$ as the reference frame, depending on which PSNR value is larger. This algorithm allows the selection of keyframe that have information similar to the intermediate frame and enhances the intermediate frame. 



\begin{figure}[t]
\centering
\includegraphics[width=0.8\textwidth]{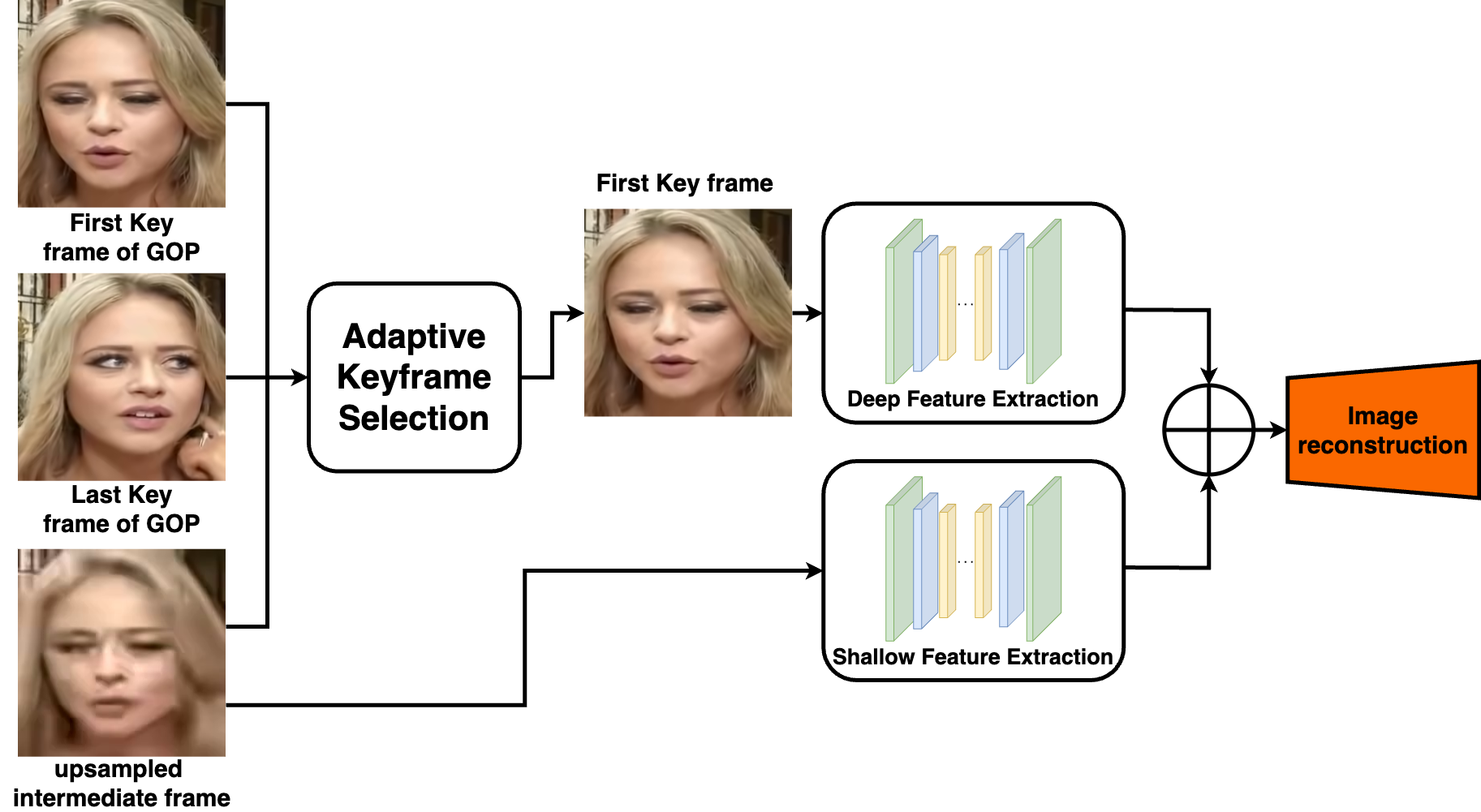}
\caption{The structure of upsampled intermediate frame enhancement.}\label{fig:fig3}
\end{figure}

To enhance the upsampled intermediate frames, we employed the SwinIR framework \cite{liang2021swinir}. 
SwinIR consists of three modules: shallow feature extraction, deep feature extraction and high-quality image reconstruction, and are intended for image restoration.
SwinIR has various tasks: image super-resolution, image denoising, and JPEG compression artifact reduction.
The upsampled intermediate frame and the selected keyframe are taken as input. Then, the shallow features extracted from the upsampled intermediate frame and the deep features sampled from the keyframe are aggregated. Finally, the image enhancement is conducted as shown in Figure\ref{fig:fig3}.

\subsection{Bidirectional Reference-Guided Video Reconstruction (BRG-VRec)}
\label{sec:BRG-VRec}
 
As we have transmitted the first and last frames in high quality, they are directly utilized as the output video, while only the intermediate frames are reconstructed. 
For intermediate frame reconstruction, as in the case of upsampled intermediate frame enhancement, the keyframe is selected from two high-quality keyframes which has similar information to the intermediate frame as shown in Algorithm\ref{alg1}.
We estimate the keypoints from the selected keyframe using the U-Net as in the case of intermediate keypoints estimation. 
The selected keyframe, its keypoints, and intermediate keypoints are used to generate the optical flow and the occlusion map. The dense optical flow captures object movement between the keyframe and the target frame, aligning the keyframe's feature map with the object's pose in the target frame. The occlusion map masks regions of the feature map for inpainting.
The animation frames $\boldsymbol{animations}$ are generated in a similar process to \cite{Siarohin_2019_NeurIPS}.

If $key_1$ or $key_2$ is selected, the animation frame is generated using the corresponding keyframe ($key_1$ or $key_2$), the keyframe's keypoints ($key\_kp_1$ or $key\_kp_2$), and $inter\_kp_i$ from the intermediate frame.



After the animation frames are generated, the feature of the animation frame $animation \in animations$ and the feature of the auxiliary frame $enh\_aux \in enh\_auxs$ are concatenated. Then the pretrained decoder $D_F$ \cite{konuko2022hybrid} reconstructs the output video frame ${DECI}$.


\subsection{Loss function and training}
We initialize the intermediate frame enhancement with pretrained models from SwinIR. To enhance the low quality intermediate frames, we used the Charbonnier loss and ${\varepsilon }$ is a constant that is set to $10^{-3}$.
\begin{center}
$L_{enhancement}=\sqrt{{||enh\_aux-I||}^2+\varepsilon}$
\end{center}
The reconstruction network is also trained with a loss term following HDAC \cite{konuko2022hybrid}.

\section{EXPERIMENTS AND RESULTS}
\subsection{Datasets and metrics}
As with \cite{konuko2021ultra,konuko2022hybrid,konuko2023predictive}, we employed the VoxCeleb dataset \cite{nagrani2017voxceleb}. It is a large audio-visual dataset of human speech extracted from YouTube videos. 
We utilized videos with a resolution of 256x256 pixels. For evaluating the results, we randomly selected 45 sequences.
As described in \cite{chen2022beyond,konuko2023predictive}, traditional objective quality assessment methods such as PSNR and SSIM are designed for pixel-level distortion calculations and are not suitable for quality assessment of deep learning-based compression. Visual quality measures in learning-based image compression, such as DISTS \cite{ding2020image} and LPIPS \cite{zhang2018unreasonable}, quantify the similarity between the reconstructed image and the original image through learning-based feature correlation and can better assess the quality of the generated image. These assessments indicate that a lower score corresponds to better quality.

\begin{table}[H]
    \caption{Comparison of the coding performance with SOTAs}
    \centering
    \begin{adjustbox}{width=\textwidth}
        \begin{tabular}{c|c|c|c|c}
          \hline
          \textbf{} & vs. HEVC \cite{6316136} & vs. VVC \cite{bross2021overview} & vs. HDAC \cite{konuko2022hybrid} & vs. RDAC \cite{konuko2023predictive} \\
          \textbf{} & BD quality / BD rate & BD quality / BD rate & BD quality / BD rate & BD quality / BD rate\\
          \hline
          LPIPS\_Alex↓ & -3.134 / -31.110 & -4.198 / -29.078 & -0.465 / -24.746 & -1.966 / -52.018 \\
          LPIPS\_VGG↓ & -6.218 / -53.410 & -7.363 / -48.266 & -0.751 / -27.795 & -3.311 / -53.886 \\
          DISTS↓ & -2.032 / -30.223 & -2.677 / -28.381 & -0.365 / -22.862  & -2.799 / -66.859 \\
          \hline
        \end{tabular}
    \end{adjustbox}    
    \label{tb:BD-rate}
\end{table}


\begin{figure}[H]
\centering
\includegraphics[width=0.9\textwidth]{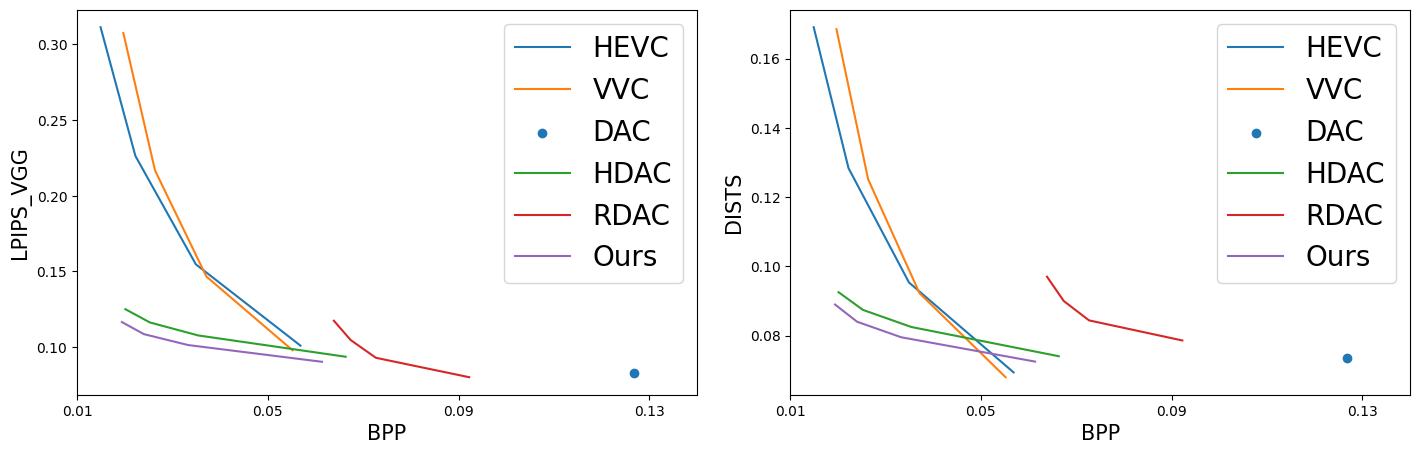}
\caption{RD performance comparison}
\label{fig:fig4}
\end{figure}

\subsection{Comparison with state-of-the-art video codecs}
To compare the performance at different bitrates, the GOP size is changed to affect the average reconstruction quality of the video sequence. 120 frames of each video are tested with GOP sizes 5, 10, 15 and 20. The quantization parameters (QP) are set to 30 for keyframes and 45 for auxiliary frames.
HEVC \cite{6316136} and VVC \cite{bross2021overview} were tested under a low-delay configuration with high QP settings to obtain a smaller bitrate.
We also compare with the animation based video codecs: DAC \cite{konuko2021ultra}, HDAC \cite{konuko2022hybrid}, and RDAC \cite{konuko2023predictive}. 
For a fair comparison, both these animation based codecs and ours, keyframes are compressed with a QP setting of 30 using VVC. 
In addition, since these codecs have various parameters for testing, we used the original ones for those parameters and obtained different bitrate results by changing the GOP.

Figure \ref{fig:fig4} provides the rate-distortion performance of the proposed codec in comparisons with conventional video codecs in terms of both LPIPS and DISTS. The rate-distortion performance is averaged across all sequences in the test data. 
It can be observed that our codec obtains higher performance at low bitrate compared to HEVC and VVC. It also achieved better results most of the time compared to other animation based video codecs.

Table \ref{tb:BD-rate} shows the bitrate savings in LPIPS\_Alex, LPIPS\_VGG, and DISTS compared to state of the art codecs (SOTAs).
Our codec delivers a 29\% bitrate reduction in LPIPS\_Alex, 48\% in LPIPS\_VGG, and 28\% in DISTS compared to the video coding standard VVC.
Furthermore, our codec achieves a 52\% bitrate reduction in LPIPS\_Alex, 53\% in LPIPS\_VGG, and 66\% in DISTS compared to the latest animation based video codec RDAC \cite{konuko2023predictive}. 

\begin{figure}[t]
\centering
\includegraphics[width=\textwidth]{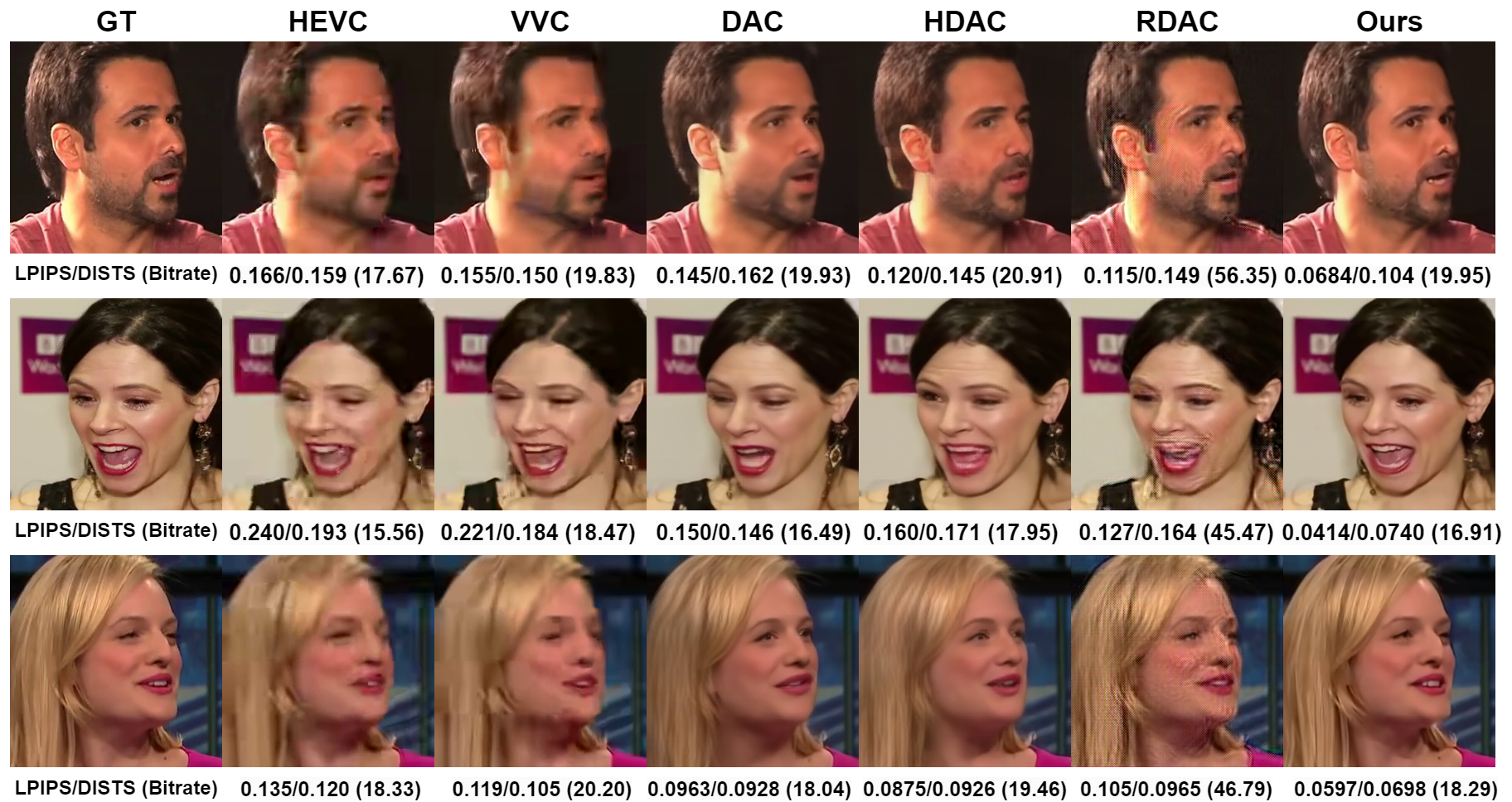}
\caption{Comparison of visual quality with SOTAs. GT stands for Ground Truth frame.}\label{fig:fig5}
\end{figure}

As shown in Figure \ref{fig:fig5}, at the similar bitrate, our approach demonstrates superior subjective quality compared to other video compression model. Additionally, it effectively maintains lower LPIPS\_Alex and DISTS scores compared to other methods. These results indicate that the bidirectional learned video reconstruction method utilizing past and future keyframes achieves better frame generation accuracy when compared to the unidirectional video reconstruction method using a single keyframe.

\begin{figure}[H]
\centering
\includegraphics[width=\textwidth]
{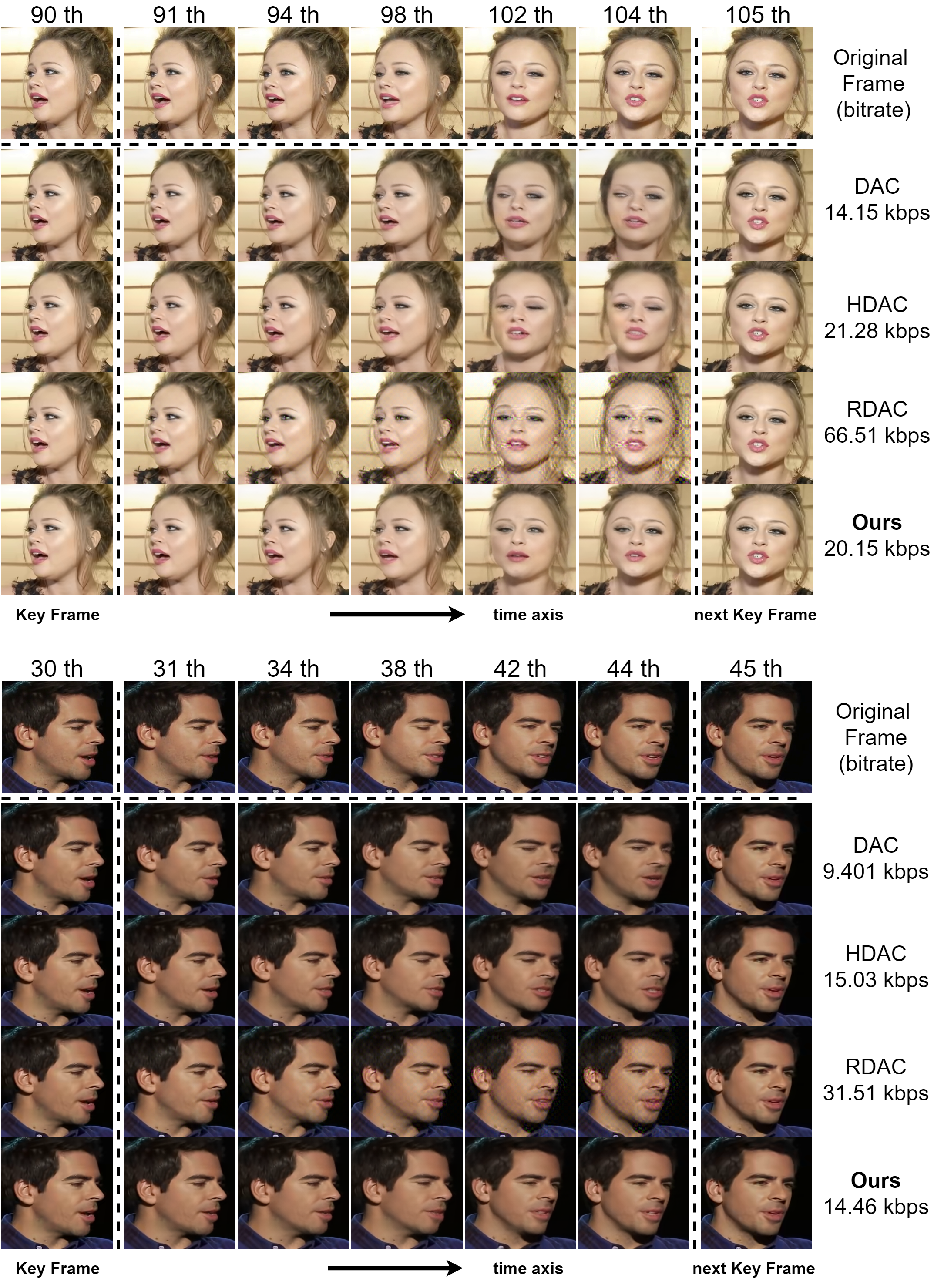}
\caption{Changes in reconstructed frame quality over time (GOP=15)}
\label{fig:fig6}
\end{figure}

Figure \ref{fig:fig6} shows the changes over time of the reconstructed frame quality when the GOP is set to 15.
Both DAC \cite{konuko2021ultra} and HDAC \cite{konuko2022hybrid} generate frames using a single keyframe, with DAC relying on keypoints for the target frame and HDAC additionally compositing auxiliary frames onto the animation frame generated similarly to DAC.
These methods can accurately reconstruct the target frame in the vicinity of the keyframe, but less accurate as time progresses as shown in Figure \ref{fig:fig6}.
The latest RDAC \cite{konuko2023predictive} reconstructs frames by combining the residuals of the animation frame as DAC and the target frame, but as time progresses, the residuals do not synthesize well and appear as noise. Also, as shown in the upper part of Figure \ref{fig:fig6}, if the motion between frames is large, the bitrate is larger because more bits of the residuals are sent to the decoder.
Our bidirectional frame reconstruction with adaptive keyframe selection can suppress the degradation of reconstruction accuracy as time progresses. This has been shown to reduce the loss of temporal correlation.

\section{CONCLUSION}
In this paper, we propose a novel bidirectional learned video codec that combines the image animation method with a downsampling based video coding. 
Our approach leverages two keyframes while enhancing downsampled/upsampled auxiliary frames. 
The reconstruction of intermediate frames is achieved using two high-quality keyframes, keypoints, and auxiliary frames. 
This method effectively addresses the loss of temporal correlation problem by utilizing two keyframes and adaptive keyframe selection, allowing for reconstruction with information closely aligned to the target intermediate frame.
Our results demonstrate significant advantages over existing compression standards such as HEVC and VVC at low bitrates. 
Also, compared to existing unidirectional facial animation based video codecs, our bidirectional approach can reduce distortion in facial regions.
It also maintains the same quality as current codecs but at a lower bitrate.

\section{REFERENCES}
\bibliographystyle{IEEEbib.bst}
\bibliography{refs.bib}

\end{document}